\documentclass[aps,prd,10pt,twocolumn,nofootinbib]{revtex4-1}
\usepackage{graphicx, epsfig}
\usepackage{amsmath,amssymb,amsfonts,dsfont,mathrsfs,amsthm,mathtools,physics}
\usepackage[dvipsnames]{xcolor}
\usepackage[unicode]{hyperref}
\hypersetup{colorlinks=true, citecolor=MidnightBlue,
            linkcolor=MidnightBlue, urlcolor=MidnightBlue, linktocpage=true}

\usepackage{array}
\usepackage{hyperref}
\usepackage{lipsum}
\usepackage{microtype}

\begin{document}

\title{Gravitational Cheshire effect from Yang--Mills instantons}

\author{Borja \surname{Diez}}
\email{borja.diez@cinvestav.mx}
\affiliation{Departamento de F\'{\i}sica, Cinvestav, Av.~IPN 2508, 07360, CDMX, M\'exico}

\begin{abstract}
We show that the dimensional reduction of four-dimensional Yang--Mills instantons naturally gives rise to nontrivial matter configurations in three dimensions exhibiting an unusual mechanism of gravitational screening. Starting from Euclidean Yang--Mills theory, we reduce (anti-)self-dual gauge configurations along a compact direction generated by a $U(1)$ isometry of the instanton background. Under this reduction, the gauge connection decomposes into a three-dimensional Yang--Mills field and an adjoint Higgs field, while the metric gives rise, as usual, to a dilaton and an Abelian gauge connection. For instantons on the Taub–NUT background, the reduced fields organize themselves into two mutually screening pairs: the backreaction of the Abelian sector cancels internally, and the same occurs independently in the non-Abelian sector. As a result, the full matter content supports a composite stealth configuration on flat three-dimensional space. By contrast, the dimensional reduction of instantons on the Eguchi–Hanson background yields a non-Abelian matter sector whose gravitational backreaction cancels, while the Abelian sector remains gravitationally active and therefore deforms the three-dimensional background geometry. These examples reveal a direct connection between four-dimensional (anti-)self-duality and three-dimensional BPS-type relations, and illustrate how instantonic sectors that are gravitationally invisible in four dimensions can descend into either fully composite stealth configurations or partially stealth matter configurations after dimensional reduction.
\end{abstract}

\maketitle

\section{Introduction}\label{sec:intro}
General Relativity provides the standard classical description of gravitational interactions, identifying gravity with the geometry of spacetime and relating curvature to the presence of matter through Einstein's field equations. From this perspective, one might naively expect that a nontrivial matter configuration should necessarily deform the background geometry. However, this expectation is not always fulfilled. A remarkable counterexample was exhibited in Ref.~\cite{Ayon-Beato:2005yoq}, where General Relativity coupled to a nonminimally interacting scalar field was shown to admit nontrivial matter configurations propagating on flat spacetime without producing any gravitational backreaction. In such configurations, the Einstein tensor and the matter energy-momentum tensor vanish independently, rather than only after being equated through the gravitational field equations. This phenomenon was dubbed the \emph{gravitational Cheshire effect}~\cite{Ayon-Beato:2005yoq}, in analogy with the Cheshire Cat in Lewis Carroll's \emph{Alice's Adventures in Wonderland}, which disappears while leaving only its grin behind.

The underlying mechanism relies on a nontrivial cancellation among different contributions to the matter energy-momentum tensor, typically involving minimal and nonminimal couplings, such that the total energy-momentum tensor vanishes identically on a given background geometry. These configurations therefore represent nontrivial stationary points of the matter action not only with respect to the matter fields themselves, but also with respect to variations of the gravitational background. They constitute particular examples of gravitational \emph{stealth} configurations. Moreover, despite being gravitationally undetectable at the level of the background geometry, these configurations can leave nontrivial thermodynamic imprints. In particular, their presence can screen the black hole mass and modify the entropy~\cite{Erices:2024iah}. One of the earliest examples of this phenomenon in a black-hole background was found for the static Bañados--Teitelboim--Zanelli (BTZ) black hole~\cite{Banados:1992wn} in three dimensions, nonminimally coupled to a scalar field~\cite{Ayon-Beato:2004nzi}. Similar configurations were subsequently constructed in more exotic backgrounds, including Lifshitz spacetimes~\cite{Ayon-Beato:2015qfa} and cosmological settings~\cite{Banerjee:2006pr,Maeda:2012tu,Ayon-Beato:2013bsa,Ayon-Beato:2015mxf}. Within scalar-tensor theories of the Horndeski family~\cite{Horndeski:1974wa}, stealth configurations have also been reported on Schwarzschild backgrounds~\cite{Babichev:2013cya,Minamitsuji:2018vuw,Giardino:2023qlu}, while extensions to beyond-Horndeski theories~\cite{Gleyzes:2014dya} have produced analogous solutions with nonconstant kinetic terms~\cite{Bakopoulos:2023fmv} and on rotating backgrounds~\cite{Charmousis:2019vnf}. More recently, related stealth configurations have also been found in theories involving non-Noetherian conformal scalar sectors~\cite{Ayon-Beato:2024xgp,Fernandes:2021dsb,Ayon-Beato:2023bzp}. 

A different realization of the same general idea arises when several matter fields interact with one another in such a way that their individual gravitational backreactions mutually cancel. In that case, the separate energy-momentum tensors need not vanish individually, whereas their sum does. Such configurations may be regarded as \emph{composite stealths}. An explicit three-dimensional example was presented in Ref.~\cite{Cardenas:2014kaa}, where the energy-momentum tensors associated with an Abelian gauge field and a conformally coupled scalar field cancel each other, leaving the geometry identical to that of the uncharged static BTZ black hole despite the presence of nontrivial matter fields. Another example of a composite stealth configuration was constructed in Ref.~\cite{Flores-Alfonso:2023fkd}, where a similar mechanism occurs in both Minkowski and AdS spacetimes. 

Interestingly, one of the earliest and most natural realizations of gravitational invisibility is already present in Yang--Mills theory. Self-gravitating Yang--Mills systems are notoriously nonlinear and are therefore often studied numerically. Nevertheless, the non-Abelian structure of the theory allows for distinguished classes of exact configurations. One important example is provided by merons~\cite{deAlfaro:1976qet}, genuinely non-Abelian gauge configurations that are locally proportional to a pure-gauge connection but are not themselves pure gauge because of the nonlinear commutator term in the field strength. It has been shown that meronic configurations can consistently gravitate~\cite{Canfora:2012ap,Aviles:2026fsf,Canfora:2023bug,Canfora:2022nso,Corral:2024xfv,Flores-Alfonso:2024gag}, and they have recently been extended to theories beyond standard Yang--Mills~\cite{Diez:2026bmp}.

Another distinguished class is provided by Yang--Mills instantons, namely finite-action Euclidean configurations satisfying the (anti-)self-duality condition on the gauge field strength. Such configurations automatically solve the second-order Yang--Mills equations and minimize the Euclidean action within a fixed homotopy class. They interpolate between topologically distinct yet degenerate vacuum sectors, thereby revealing the nontrivial periodic vacuum structure of Yang--Mills theory~\cite{Jackiw:1976pf,Callan:1976je}. The prototypical example is the Belavin--Polyakov--Schwartz--Tyupkin instanton on $\mathbb{R}^4$~\cite{Belavin:1975fg}, which has subsequently been generalized to other constant-curvature backgrounds~\cite{Callan:1989em} and to theories beyond standard Yang--Mills~\cite{Canfora:2025gwm}. From the gravitational point of view, (anti-)self-dual Yang--Mills fields are especially remarkable because their energy-momentum tensor vanishes identically. They therefore provide a particularly simple realization of a stealth matter sector in four-dimensional Euclidean space. Physically, Yang--Mills instantons describe nonperturbative saddle points interpolating between topologically inequivalent gauge vacua and provide the leading semiclassical contribution to tunneling amplitudes in the path-integral formulation of non-Abelian gauge theories.

A natural framework in which such configurations can be related to lower-dimensional gravitating systems is provided by Kaluza--Klein (KK) dimensional reduction. In the original construction of Kaluza~\cite{Kaluza:1921tu} and Klein~\cite{Klein:1926tv}, five-dimensional vacuum Einstein gravity was shown, after compactification along a circular direction, to give rise to four-dimensional gravity coupled to an Abelian gauge field and a scalar dilaton. Dilatonic couplings subsequently became ubiquitous in effective gravitational theories, particularly in low-energy limits of supergravity and string-inspired models~\cite{Gibbons:1987ps,Garfinkle:1990qj}. In non-Abelian systems, the dimensional reduction of Yang--Mills fields naturally produces Higgs fields in the adjoint representation. For instance, monopole solutions of $SU(2)$ Yang--Mills--Higgs theory, such as the 't Hooft--Polyakov monopole~\cite{tHooft:1974kcl,Polyakov:1974ek} and its higher-winding generalizations~\cite{Ward:1981jb,Rebbi:1980yi,Forgacs:1980ym}, have been extensively studied in the presence of a massless dilaton~\cite{Forgacs:1995zq,Brihaye:2001gg}. These systems display several similarities with gravitating monopoles in Einstein--Yang--Mills--Higgs theory~\cite{Lee:1991vy,Breitenlohner:1991aa,Breitenlohner:1994di}, including the possibility that the dilaton mediates an effective attraction between magnetically charged objects and allows for bound multimonopole configurations.

One of the most celebrated consequences of KK reduction is the appearance of lower-dimensional solitonic configurations whose charges have a purely geometrical origin. The canonical example is the Gross--Perry--Sorkin (GPS) monopole~\cite{Gross:1983hb,Sorkin:1983ns}, commonly referred to as the KK monopole. It arises from the dimensional reduction of the Taub--Newman--Tamburino--Unti (Taub--NUT) gravitational instanton~\cite{Taub:1950ez,Newman:1963yy} after oxidation by an additional timelike direction. From the higher-dimensional perspective, the geometry is a regular Ricci-flat solution, whereas after reduction along the compact $U(1)$ fiber it is described by a magnetically charged configuration of the Einstein--Maxwell--dilaton system. Higher-dimensional generalizations of this construction have also been explored~\cite{Mann:2005gk,AvilesDiez}. In a related direction, Volkov showed that five-dimensional Einstein--Yang--Mills theory reduces to a four-dimensional Einstein--Yang--Mills--Higgs system with nontrivial dilatonic couplings fixed by the higher-dimensional origin of the theory~\cite{Volkov:2001tb}.

In this work, we exploit this dimensional-reduction mechanism to uncover a direct relation between four-dimensional Yang--Mills instantons and lower-dimensional composite stealth configurations. Starting from Euclidean Yang--Mills instantons on gravitational instanton backgrounds, we reduce the system along a compact direction associated with a $U(1)$ isometry. The higher-dimensional gauge connection decomposes into a three-dimensional Yang--Mills field and an adjoint Higgs field, whereas the metric gives rise to the usual dilaton and Abelian gauge field. Remarkably, the gravitational invisibility of the parent (anti-)self-dual Yang--Mills configuration descends into nontrivial screening relations among the reduced matter sectors.


The remainder of this manuscript is organized as follows. In Sec.~\ref{sec:EYM theory}, we introduce four-dimensional Einstein--Yang--Mills theory with gauge group $SU(2)$ and establish the notation and conventions used throughout the paper. We then review Yang--Mills instanton solutions supported on gravitational-instanton backgrounds, focusing in particular on the (anti-)self-dual Taub--NUT and Eguchi--Hanson geometries. In Sec.~\ref{sec:the model}, we perform the dimensional reduction of the Einstein--Yang--Mills system, obtaining a three-dimensional Einstein--Yang--Mills--Higgs--dilaton theory coupled to an Abelian gauge field. In Sec.~\ref{sec:composite-flat}, we show that the reduction of the Yang--Mills instanton on Taub--NUT gives rise to a composite stealth configuration on flat Euclidean space, supported by nontrivial matter fields whose gravitational backreactions cancel pairwise. In Sec.~\ref{sec:partially composite}, we demonstrate that the reduction of the instanton on Eguchi--Hanson instead yields a partially composite stealth configuration, in the sense that only one subset of the matter fields exhibits mutual cancellation of its backreaction, while the remaining sector continues to deform the background geometry. The relevant limiting regimes of both classes of solutions are also discussed. Finally, in Sec.~\ref{sec:discussion}, we summarize our main results and comment on several possible directions for future research.

\section{Einstein--Yang--Mills theory and instantons}\label{sec:EYM theory}
Our starting point is four-dimensional Euclidean Einstein--Yang--Mills theory with gauge group $SU(2)$, whose dynamics is governed by the action
\begin{equation}\label{S-EYM}
	\hat{S}[g_{MN},A_M]
	=
	\frac{1}{2}
	\int \dd^4x\sqrt{|\hat g|}
	\left(
	\hat R+\Tr F_{MN}F^{MN}
	\right),
\end{equation}
where $\hat R=\hat R_{MN}\hat g^{MN}$ is the four-dimensional Ricci scalar. Throughout this work we adopt units in which $8\pi \hat G_N=1$, with $\hat G_N$ denoting the four-dimensional Newton constant. The non-Abelian gauge potential is taken in the adjoint representation of $SU(2)$, that is, $A_M=A_M^a t_a$, with associated field strength
\begin{equation}
	F_{MN}
	=
	F^a_{MN}t_a
	=
	\partial_M A_N-\partial_N A_M+[A_M,A_N]\,.
\end{equation}
The latter satisfies the non-Abelian Bianchi identity
\begin{equation}
	\nabla_M\widetilde F^{MN}
	+
	[A_M,\widetilde F^{MN}]
	=
	0,
\end{equation}
where $\widetilde F_{MN}=\frac{1}{2}\epsilon_{MNPQ}F^{PQ}$ is the Hodge dual of the Yang--Mills field strength, with $\epsilon_{MNPQ}$ denoting the four-dimensional Levi-Civita tensor. We choose the anti-Hermitian generators to be $t_a=-\frac{i}{2}\tau_a$, where $\tau_a$ are the standard Pauli matrices satisfying
\begin{equation}
	[\tau_a,\tau_b]
	=
	2i\epsilon_{abc}\tau_c,
	\qquad
	\{\tau_a,\tau_b\}
	=
	2\delta_{ab}\mathbb{I}.
\end{equation}
Lowercase Latin indices $a,b,\ldots$ denote internal $SU(2)$ indices, whereas uppercase Latin indices $M,N,\ldots$ label four-dimensional spacetime coordinates. We shall decompose the latter as $x^M=\{\psi,x^\mu\}$, where Greek indices $\mu,\nu,\ldots$ refer to the three-dimensional space obtained after dimensional reduction. Hatted quantities will systematically denote four-dimensional objects, while unhatted ones will refer to their three-dimensional counterparts.

The equations of motion following from the action~\eqref{S-EYM} are
\begin{subequations}\label{eom-EYM}
	\begin{align}
		R_{MN}-\frac{1}{2}\hat R g_{MN}&=T_{MN},
		\label{eom-YM}\\
		\nabla_MF^{MN}+[A_M,F^{MN}]&=0,
		\label{eom-ym}
	\end{align}
\end{subequations}
where the Yang--Mills energy-momentum tensor is defined as
\begin{equation}
	T_{MN}
	=
	-2\Tr\left(
	F_{MP}F_N{}^{P}
	-\frac{1}{4} g_{MN}F_{PQ}F^{PQ}
	\right).
\end{equation}

Owing to their non-Abelian character, the Yang--Mills equations form a highly nonlinear system of second-order differential equations and are notoriously difficult to solve in generic curved backgrounds. A distinguished class of configurations is obtained by imposing the first-order (anti-)self-duality condition
\begin{equation}\label{sd-F-d4}
	F_{MN}
	=
	\pm \widetilde F_{MN}\,,
\end{equation}
which, together with the Bianchi identity, Eq.~\eqref{sd-F-d4} automatically implies the Yang--Mills equations~\eqref{eom-ym}. Furthermore, (anti-)self-dual configurations saturate the standard Bogomol'nyi bound on the Euclidean Yang--Mills action within a fixed topological sector, leaving the on-shell action proportional to the second Chern--Pontryagin number, which characterizes the corresponding homotopy class.

An equally remarkable consequence of Eq.~\eqref{sd-F-d4} is that the Yang--Mills energy-momentum tensor vanishes identically. Indeed, it may be written as
\begin{equation}\label{Tmunu-Instanton}
	T_{MN}
	=
	-\Tr\left[
	(F_{MP}+\widetilde F_{MP})
	(F_{NQ}-\widetilde F_{NQ})
	\right]g^{PQ},
\end{equation}
which vanishes identically whenever the field strength is either self-dual or anti-self-dual. Consequently, Yang--Mills instantons carry nontrivial gauge and topological structure while producing no local gravitational backreaction. From this perspective, they provide a natural realization of gravitational stealth configurations in the Euclidean sector.

Yang--Mills instantons are therefore finite-action solutions of the Euclidean Yang--Mills equations satisfying the first-order condition~\eqref{sd-F-d4}. Although the latter is simpler than the original second-order field equations, finding explicit instantons on generic curved spaces remains highly nontrivial. A particularly favorable situation arises when the background geometry is itself (anti-)self-dual, namely when its Riemann tensor satisfies
\begin{equation}\label{sd-R-d4}
	 R_{MNST}
	=
	\pm\frac{1}{2}
	\epsilon_{MNPQ}
	 R^{PQ}{}_{ST}.
\end{equation}
Ricci-flat Euclidean geometries obeying Eq.~\eqref{sd-R-d4} constitute the canonical examples of gravitational instantons, so named because of their close geometric analogy with Yang--Mills instantons.

The interplay between gauge and gravitational self-duality considerably simplifies the construction of non-Abelian instantons. In Ref.~\cite{Kim:2000mg}, an algorithm was developed to generate (anti-)self-dual Yang--Mills configurations on gravitational-instanton backgrounds. The resulting solutions are generically anisotropic in the internal space, in the sense that the gauge-field profiles associated with the different generators of $SU(2)$ need not coincide. In what follows, we focus on two of the most prominent gravitational instantons, namely Taub--NUT and Eguchi--Hanson. Both belong to the broader Gibbons--Hawking class of hyper-K\"ahler geometries, which can be constructed in terms of harmonic functions on three-dimensional Euclidean space $\mathbb{R}^3$~\cite{Gibbons:1978tef,Gibbons:1987sp}. Their first-order $\alpha'$ corrections within the low-energy effective action of heterotic string theory have also been investigated recently in Ref.~\cite{Cerdeira:2026uem}.

\subsection{Instanton on the Taub--NUT background}

An exact solution of the Einstein--Yang--Mills system~\eqref{eom-EYM} is provided by an anti-self-dual Yang--Mills instanton on the Taub--NUT gravitational instanton. The corresponding gauge configuration was originally constructed by Pope and Yuille~\cite{Pope:1978kj} following the prescription introduced by Charap and Duff~\cite{Charap:1977ww}. To describe the geometry, we introduce the $SU(2)$ left-invariant one-forms
\begin{subequations}
	\begin{align}
		\sigma_1
		&=
		\sin\psi\,\dd\vartheta
		-\cos\psi\sin\vartheta\,\dd\varphi,
		\\
		\sigma_2
		&=
		\cos\psi\,\dd\vartheta
		+\sin\psi\sin\vartheta\,\dd\varphi,
		\\
		\sigma_3
		&=
		\dd\psi+\cos\vartheta\,\dd\varphi,
	\end{align}
\end{subequations}
which satisfy the Maurer--Cartan relations
\begin{equation}
	\dd\sigma_a
	+
	\frac{1}{2}
	\epsilon_{abc}\sigma_b\wedge\sigma_c
	=
	0.
\end{equation}
The coordinates $(\vartheta,\varphi,\psi)$ are Euler angles with ranges $\varphi\in [0,\pi], \varphi\in [0,2\pi)$, and $\psi\in [0,4\pi )$. In these coordinates, the anti-self-dual Taub--NUT metric is given by~\cite{Taub:1950ez,Newman:1963yy}
\begin{equation}\label{TN-d4}
\begin{split}
	\dd\hat s^2
	&=
	\hat g_{MN}\dd x^M\dd x^N
	\\
	&=
	\left(\frac{r+n}{r-n}\right)\dd r^2
	+
	4n^2
	\left(\frac{r-n}{r+n}\right)\sigma_3^2
	+
	(r^2-n^2)\dd\Omega_2^2,
\end{split}
\end{equation}
where
\begin{equation}
	\dd\Omega_2^2
	=
	\sigma_1^2+\sigma_2^2
	=
	\dd\vartheta^2
	+
	\sin^2\vartheta\,\dd\varphi^2
\end{equation}
is the metric of the unit round two-sphere. The coordinate $\psi$ parametrizes the compact $U(1)$ fiber.

The metric~\eqref{TN-d4} has an anti-self-dual Riemann tensor and covers the region $r\geq n$. At $r=n$, the $U(1)$ fiber collapses to a zero-dimensional fixed-point set, defining a \emph{nut}. Despite the degeneration of the coordinate system, the full four-dimensional geometry is completely regular there. If the nut is excluded, the manifold has topology $\mathbb{R}\times \mathbb{S}^3$, whereas including it completes the geometry to a space topologically equivalent to $\mathbb{R}^4$~\cite{Eguchi:1977iu,Hawking:1976jb}. This spacetime has also been generalized to include multiple NUT parameters~\cite{Mann:2005ra,Corral:2026tjy}.

Remarkably, the anti-self-duality equations for the Yang--Mills field can be solved exactly on this background. A one-parameter family of solutions is given by~\cite{Pope:1978kj,Yuille:1979gj}
\begin{equation}\label{TN-A-d4}
\begin{split}
    \hat{A}&=A_M\dd x^M\\
    &=\left(\frac{r-n}{r+a}\right)(\sigma_1t_1+\sigma_2t_2)+ \frac{r-n}{r+n}\left(1+\frac{2n}{r+a}\right)\sigma_3t_3\,,
\end{split}
\end{equation}
where $a$ is an integration constant satisfying $a\in[n,\infty)$. For $a=n$, the configuration reduces to the Pope--Yuille instanton~\cite{Pope:1978kj}, whereas the opposite limiting regime, namely,  $a\rightarrow\infty$ yields the corresponding dyonic configuration.

The associated non-Abelian field strength is anti-self-dual and consequently has an identically vanishing energy-momentum tensor by virtue of Eq.~\eqref{Tmunu-Instanton}. Thus, despite carrying nontrivial gauge structure, the Yang--Mills field produces no backreaction on the Taub--NUT gravitational instanton.

At the nut, $r=n$, the gauge connection becomes trivial, consistently with the collapse of the compact fiber. At large radius, on the other hand, the field approaches a pure-gauge configuration. The Taub--NUT geometry itself is asymptotically locally flat (ALF): as $r\rightarrow\infty$, its metric approaches a nontrivial $\mathbb{S}^1$ fibration of finite asymptotic radius over $\mathbb{R}^3$.

\subsection{Instanton on the Eguchi--Hanson background}
A second exact solution of the four-dimensional Einstein--Yang--Mills system~\eqref{eom-EYM} is obtained by placing a self-dual Yang--Mills instanton on the Eguchi--Hanson gravitational instanton. The corresponding line element reads
\begin{equation}\label{EH-d4}
\begin{split}
	\dd\hat s^2
	&=
	\hat g_{MN}\dd x^M\dd x^N
	\\
	&=
	\left(
	1-\frac{a^4}{r^4}
	\right)^{-1}\dd r^2
	+
	\frac{r^2}{4}
	\left(
	1-\frac{a^4}{r^4}
	\right)\sigma_3^2
	+
	\frac{r^2}{4}\dd\Omega_2^2,
\end{split}
\end{equation}
with $r\geq a$. The associated Yang--Mills instanton may be written as~\cite{Kim:2000mg}
\begin{equation}\label{EH-A-d4}
\begin{split}
	\hat{A}&=A_M\dd x^M\\
    &=\sqrt{1-\frac{a^4}{r^4}}\left(\sigma_1t_1+\sigma_2t_2\right)+\left(1+\frac{a^4}{r^4}\right)\sigma_3t_3.
\end{split}
\end{equation}

The hypersurface $r=a$ corresponds to a \emph{bolt}, namely a two-dimensional fixed-point set of the Killing vector generating the $U(1)$ action. Although the metric coefficients in Eq.~\eqref{EH-d4} appear singular at this location, the apparent divergence is merely a coordinate artifact. Regularity requires the fiber coordinate to have period $2\pi$ rather than the $4\pi$ periodicity of the covering $\mathbb{S}^3$. Consequently, constant-$r$ hypersurfaces at large radius have topology $\mathbb{S}^3/\mathbb{Z}_2$ and the full Eguchi--Hanson manifold is an asymptotically locally Euclidean space. Its Euler characteristic is $\chi=2$, and its Hirzebruch signature is $\tau=-1$. Generalizations of this geometry in the presence of a cosmological constant, Maxwell fields, and higher-curvature corrections have been explored in Refs.~\cite{Corral:2022udb,Corral:2025yvr,Eguchi:1978gw,Pedersen1985EguchiHansonMW}. The Eguchi--Hanson space has also been employed as a seed geometry for the construction of five-dimensional gravitational solitons~\cite{Clarkson:2005qx,Clarkson:2006zk,Chng:2006gh,Wong:2011aa,Durgut:2022xzw}.

At the bolt $r=a$, the non-Abelian gauge configuration becomes effectively Abelian, since the components along $t_1$ and $t_2$ vanish while only the Cartan component survives. Conversely, in the asymptotic region $r\rightarrow\infty$, the gauge field approaches a pure-gauge configuration. As in the Taub--NUT case, its self-duality guarantees that the four-dimensional Yang--Mills energy-momentum tensor vanishes identically, rendering the gauge instanton gravitationally invisible despite its nontrivial topological content.

\section{The model}\label{sec:the model}
Dilaton fields nonminimally coupled to Abelian and non-Abelian gauge sectors arise naturally in low-energy effective descriptions inspired by string theory and supergravity. In particular, exponential dilaton couplings to gauge kinetic terms constitute a characteristic feature of effective actions associated with the massless string spectrum, where gravity, scalar moduli, and Yang--Mills fields are generically intertwined~\cite{Fradkin:1985ys,Maharana:1992my,deRoo:1992zp}. Such interactions have also played a central role in the construction of dilatonic black holes and gravitational solitons, providing a natural framework in which scalar and gauge degrees of freedom can substantially modify the gravitational dynamics.

In the present case, the three-dimensional theory of interest arises directly from the dimensional reduction of the four-dimensional Einstein--Yang--Mills system introduced in the previous section. We reduce along a compact direction generated by a $U(1)$ isometry and adopt the KK decomposition, namely

\begin{equation}
    \dd\hat{s}=e^\phi g_{\mu\nu}\dd x^\mu\dd x^\nu + e^{-\phi}(\dd\psi+B_\mu\dd x^\mu)^2\,,
\end{equation}
together with the decomposition of the non-Abelian gauge connection
\begin{equation}
   \hat A=A_\mu\dd x^\mu+\Phi(\dd\psi+B_\mu\dd x^\mu)\,.
\end{equation}
Here, $B_\mu$ is the KK Abelian gauge field, $\phi$ is the dilaton, $A_\mu$ is a three-dimensional Yang--Mills connection, and $\Phi$ transforms in the adjoint representation of $SU(2)$ and therefore plays the role of a Higgs field from the lower-dimensional perspective. The coefficients appearing in the exponential factors are chosen so that, after dimensional reduction, both the Einstein--Hilbert term and the dilaton kinetic term are canonically normalized. 

After integrating over the compact direction, one obtains the three-dimensional Euclidean action
\begin{widetext}
\begin{equation}\label{S-d3}
	\begin{split}
		S[g_{\mu\nu},B_\mu,\phi, A_\mu,\Phi]&=\frac{1}{2}\int_\mathcal{M}\dd^3x\sqrt{|g|}\Bigg(R-\frac{1}{2}\nabla_\mu\phi\nabla^\mu\phi-\frac{1}{4}e^{-2\phi}H_{\mu\nu}H^{\mu\nu}\\
		&~~+e^{-\phi}\Tr[(F_{\mu\nu}+\Phi H_{\mu\nu})(F^{\mu\nu}+\Phi H^{\mu\nu})]+2e^\phi \Tr(D_\mu\Phi D^\mu\Phi)\Bigg)\,,
	\end{split}
\end{equation}
\end{widetext}
where $H_{\mu\nu}=\partial_\mu B_\nu-\partial_\nu B_\mu$ is the Abelian field strength, $F_{\mu\nu}=\partial_\mu A_\nu-\partial_\nu A_\mu +[A_\mu,A_\nu]$ is the non-Abelian Yang--Mills field strength, with $A_\mu=A_\mu^{a}t_a$ and $D_\mu\Phi=\nabla_\mu \Phi+[A_\mu,\Phi]$ denotes the gauge-covariant derivative of the adjoint Higgs field. Throughout this work, we adopt units in which both the gravitational and Yang--Mills couplings are set to unity. In addition, the overall factor arising from the integration over the compact direction is absorbed into the normalization of the three-dimensional action.

A distinctive feature of this model is the nonminimal interaction between the Abelian field strength and the adjoint scalar through the combination $F_{\mu\nu}+\Phi H_{\mu\nu}$, together with the dilaton-dependent gauge and Higgs kinetic terms. These couplings provide a nontrivial interplay between the gravitational, Abelian, and non-Abelian sectors while retaining the characteristic exponential scalar couplings commonly encountered in string-inspired effective theories.

Stationary variations of the action~\eqref{S-d3} with respect to the metric, the dilaton, the Yang--Mills connection, the Higgs field, and the Abelian gauge field yield,
\begin{widetext}
\begin{subequations}\label{eom-d3}
	\begin{align}
		R_{\mu\nu}-\frac{1}{2} Rg_{\mu\nu}&=T_{\mu\nu}^{(\phi)}+T_{\mu\nu}^{(B)}+T_{\mu\nu}^{(A)}+T_{\mu\nu}^{(\Phi)}\,,\\
		0&=\Box\phi+\frac{1}{2}e^{-2\phi}H_{\mu\nu}H^{\mu\nu}-e^{-\phi}\Tr[(F_{\mu\nu}+\Phi H_{\mu\nu})(F^{\mu\nu}+\Phi H^{\mu\nu})]+2e^\phi\Tr( D_\mu\Phi D^\mu\Phi)\,,\\
        0&=D_\mu[e^{-\phi}(F^{\mu\nu}+\Phi H^{\mu\nu})]-e^\phi [\Phi,D^\nu \Phi]\,,\\
        0&=D_\mu(e^\phi D^\mu\Phi)-\frac{1}{2}e^{-\phi}H_{\mu\nu}(F^{\mu\nu}+\Phi H^{\mu\nu})\,,\\
        0&=\nabla_\mu\left\{e^{-2\phi}H^{\mu\nu}-4e^{-\phi}\Tr[\Phi(F^{\mu\nu}+\Phi H^{\mu\nu})]\right\}\,,
	\end{align}
\end{subequations}
\end{widetext}
respectively, where the corresponding contributions to the total energy-momentum tensor read
\newpage
\begin{widetext}
    \begin{subequations}
        \begin{align}
            T_{\mu\nu}^{(\phi)}&=\frac{1}{2}\left(\nabla_\mu\phi\nabla_\nu\phi-\frac{1}{2}g_{\mu\nu}(\nabla\phi)^2\right)\,,\\
            T_{\mu\nu}^{(B)}&=\frac{1}{2}e^{-2\phi}\left(H_{\mu\lambda}H_\nu^{~\lambda}-\frac{1}{4}g_{\mu\nu}H_{\lambda\rho}H^{\lambda\rho}\right)\,,\\
            T_{\mu\nu}^{(A)}&=-2e^{-\phi}\Tr\left[(F_{\mu\lambda}+\Phi H_{\mu\lambda})(F_\nu^{~\lambda}+\Phi H_\nu^{~\lambda})-\frac{1}{4}g_{\mu\nu}(F_{\alpha\beta}+\Phi H_{\alpha\beta})(F^{\alpha\beta}+\Phi H^{\alpha\beta})\right]\,,\\
            T_{\mu\nu}^{(\Phi)}&=-2e^\phi\Tr\left(D_\mu\Phi D_\nu\Phi-\frac{1}{2}g_{\mu\nu}D_\lambda\Phi D^\lambda\Phi\right)\,.
        \end{align}
    \end{subequations}
\end{widetext}
The particular structure of these energy-momentum tensors will be crucial in what follows. As we shall see, for dimensional reductions of four-dimensional (anti-)self-dual Yang--Mills configurations, the first-order relations inherited from four-dimensional self-duality enforce nontrivial cancellations among the lower-dimensional matter sectors. In the Taub--NUT case, these cancellations occur independently in both the Abelian and non-Abelian sectors, yielding a fully composite stealth configuration, whereas in the Eguchi--Hanson case only the Yang--Mills--Higgs sector remains gravitationally screened.

\section{Composite Cheshire effect on flat space}\label{sec:composite-flat}
Despite the high degree of nonlinearity of the field equations~\eqref{eom-d3}, the system admits a remarkably simple exact solution obtained from the dimensional reduction of the Taub--NUT configuration given in Eqs.~\eqref{TN-d4} and~\eqref{TN-A-d4}. After suitable redefinitions of the radial coordinate and the parameters of the solution, the reduced fields can be written as
\begin{subequations}
    \begin{align}
        \dd s^2&=\dd r^2+r^2\dd\Omega^2\,,\quad \phi=\log\left(1+\frac{p}{r}\right)\,,\\
        B&=p\cos\vartheta\dd\varphi\,,\quad ~~\Phi=-\frac{(q+p)}{(2r+q+p)(r+p)}t_3\,,\\
        A&= -\frac{2r}{2r+p+q}\left(\sin\vartheta\dd\varphi t_1-\dd\vartheta t_2\right)+\cos\vartheta\dd\varphi t_3\,,
    \end{align}
\end{subequations}
where $p$ and $q$ are integration constants. The reduced metric is precisely three-dimensional Euclidean space, even though the scalar, Abelian, Yang--Mills, and Higgs fields are all nontrivial. This occurs because the corresponding matter sectors screen each other's gravitational backreaction pairwise. More precisely, the individual energy-momentum tensors satisfy
\begin{equation}
    T_{\mu\nu}^{(\phi)}=-T_{\mu\nu}^{(B)}\quad \text{and} \quad   T_{\mu\nu}^{(A)}=-T_{\mu\nu}^{(\Phi)}\,.
\end{equation}
Thus, the dilaton and the Abelian KK gauge field constitute one mutually screening pair, while the Yang--Mills and adjoint Higgs fields form a second one. Consequently, the total energy-momentum tensor vanishes identically, despite the fact that none of the individual matter sectors is trivial. The resulting configuration therefore provides an explicit realization of a \emph{composite gravitational stealth}, or equivalently a \emph{composite gravitational Cheshire effect}, in which nontrivial matter fields are present while their total gravitational imprint completely disappears.

The origin of this cancellation can be traced directly to the four-dimensional (anti-)self-duality condition. After dimensional reduction, the latter descends to the first-order relations
\begin{equation}
\begin{split}
	F_{\mu\nu}&=-e^\phi\epsilon_{\mu\nu\lambda}(\Phi\nabla^\lambda\phi+D^\lambda\Phi)\,,\\ H_{\mu\nu}&=e^\phi \epsilon_{\mu\nu\lambda}\nabla^\lambda\phi\,,
\end{split}
\end{equation}
which provide a set of first-order relations coupling the Yang--Mills connection, the adjoint Higgs field, the dilaton, and the Abelian gauge field. Their structure is reminiscent of the Bogomol'nyi--Prasad--Sommerfield (BPS) equations that arise in the theory of monopoles and solitons~\cite{Bogomolny:1975de,Prasad:1975kr} (see, for instance, Ref.~\cite{Weinberg:2012pjx}). In the present construction, however, these relations are not imposed independently in three dimensions, but are inherited from the (anti-)self-duality of the parent four-dimensional Yang--Mills instanton.

The Abelian gauge field is locally the potential of a Dirac monopole~\cite{Dirac:1931kp}, with field strength
\begin{equation}
    H=-p\sin\vartheta\dd\vartheta\wedge\dd\varphi\,.
\end{equation}

The parameter $p$ also controls the leading asymptotic behavior of the dilaton, which at large radial distance behaves as
\begin{equation}
    \phi\simeq\frac{p}{r}-\frac{p^2}{2r^2}+\mathcal{O}(r^{-3})\,.
\end{equation}
Thus, $p$ simultaneously determines the Abelian magnetic charge and the leading scalar falloff. The adjoint Higgs field instead behaves as
\begin{equation}
    \Phi
    \simeq
    -\frac{p+q}{2r^2}\,t_3
    +
    \mathcal{O}(r^{-3})\,,
\end{equation}
and therefore decays faster than the dilaton. The non-Abelian gauge connection approaches a pure-gauge configuration at infinity,
\begin{equation}
    A
    \simeq
    -\sin\vartheta\,\dd\varphi\,t_1
    +\dd\vartheta\,t_2
    +\cos\vartheta\,\dd\varphi\,t_3
    +\mathcal O(r^{-1})\,,
\end{equation}
in agreement with its instantonic four-dimensional origin.

In the limit $p=0$, both the dilaton and the Abelian Maxwell field become trivial, while the Yang--Mills and Higgs sectors remain nonvanishing. The surviving matter fields are given by\begin{equation}
\begin{split}
    \Phi&=-\frac{q}{r(2r+q)}t_3\,,\\
     A&=-\frac{2r}{2r+q}(\sin\vartheta\dd\varphi t_1-\dd\vartheta t_2)+\cos\vartheta\dd\varphi t_3\,.
\end{split}
\end{equation}
In this limit, the configuration provides a particularly simple realization of a composite stealth solution of the three-dimensional Einstein--Yang--Mills--Higgs system on flat Euclidean space. The Yang--Mills and adjoint Higgs fields remain individually nontrivial, while their energy-momentum tensors cancel each other, leaving the background geometry completely unaffected.

On the other hand, in the limit $q=-p$, the adjoint Higgs field vanishes and the Yang--Mills connection becomes pure gauge. Consequently, only the Abelian sector survives. The resulting configuration can therefore be interpreted as the the three-dimensional counterpart of the Abelian GPS monopole~\cite{Gross:1983hb,Sorkin:1983ns}, with the non-Abelian degrees of freedom becoming dynamically trivial in this limit.

Another interesting limiting regime is $q\to\infty$. In this case, the Yang--Mills connection asymptotically reduces to
\begin{equation}
    A=\cos\vartheta\,\dd\varphi\,t_3\,,
\end{equation}
while the Higgs field becomes
\begin{equation}
    \Phi=-\frac{1}{r+p}\,t_3\,.
\end{equation}
Thus, the entire matter sector effectively Abelianizes: all surviving fields are aligned along the same Cartan generator and the genuinely non-Abelian interactions disappear. The resulting configuration therefore consists solely of nontrivial Abelian fields whose combined energy-momentum tensor vanishes, so that the background geometry remains completely undeformed.

This solution therefore illustrates explicitly how a four-dimensional Yang--Mills instanton that is gravitationally invisible due to (anti-)self-duality can descend, after KK reduction, into several individually gravitating three-dimensional matter fields whose backreactions cancel only collectively. In this sense, the four-dimensional stealth sector is resolved into a pairwise composite stealth structure in the reduced theory.

%
%

\section{Partially composite stealth configuration}\label{sec:partially composite}
On the other hand, the dimensional reduction of the Yang--Mills instanton supported on the Eguchi--Hanson gravitational instanton yields the three-dimensional configuration
\begin{subequations}
    \begin{align}
        \dd s^2&=\dd r^2+\left(r^2-a^2\right)\dd\Omega^2\,,\quad \phi=-\log\left(\frac{r^2-a^2}{ar}\right)\,,\\
        B&=a\cos\vartheta\dd\varphi\,,\quad ~~\Phi=\frac{a}{r^2} t_3\,,\\
        A&= -\sqrt{1-\frac{a^2}{r^2}}\left(\sin\vartheta\dd\varphi t_1-\dd\vartheta t_2\right)+\cos\vartheta\dd\varphi t_3\,.
    \end{align}
\end{subequations}
In this case, the non-Abelian contributions to the energy-momentum tensor cancel each other, while the dilaton and Abelian KK sectors remain gravitationally active. Consequently, the total energy-momentum tensor no longer vanishes, and the reduced geometry is no longer flat. The resulting configuration therefore provides a natural example of a partially composite stealth system: the Yang--Mills--Higgs sector is gravitationally screened, whereas the Abelian sector continues backreacting on the geometry. We shall refer to such configurations as \textit{partially composite stealths}.

The three-dimensional spacetime possesses a curvature singularity at $r=a$. In particular, the Ricci scalar and the Kretschmann invariant are given by
\begin{equation}\label{curvature-EH-d3}
	R=\frac{2a^2}{(r^2-a^2)^2}\,,\quad R_{\mu\nu}^{\lambda\rho}R_{\lambda\rho}^{\mu\nu}=\frac{12a^4}{(r^2-a^2)^4}\,.
\end{equation}
Thus, both invariants diverge at $r=a$, showing that the endpoint of the reduced geometry is genuinely singular from the three-dimensional point of view.

In the asymptotic region $r\to\infty$, the dilaton behaves as
\begin{equation}
    \phi\simeq -\log(\frac{r}{a})+\frac{a^2}{r^2}+\mathcal{O}(r^{-4})\,,
\end{equation}
and therefore diverges logarithmically. Meanwhile, the Higgs field decays as
\begin{equation}
    \Phi\simeq \frac{a}{r^2}t_3\,,
\end{equation}
whereas the Yang--Mills connection approaches a pure-gauge configuration. Hence, the non-Abelian sector becomes physically trivial at infinity, while the Abelian sector remains nontrivial and is responsible for the residual gravitational backreaction.

Near the endpoint $r=a$, the two-spheres shrink to zero size and the curvature diverges, as is manifest from Eq.~\eqref{curvature-EH-d3}. This behavior originates from the degeneration of the KK circle at the Eguchi--Hanson bolt in four dimensions. In contrast with the regular four-dimensional parent geometry, the lower-dimensional quotient develops a genuine singularity at the corresponding fixed-point set. At the same time, the non-Abelian sector effectively Abelianizes, since
\begin{equation}
    A=\cos\vartheta\dd\varphi t_3\quad \text{and}\quad \Phi=\frac{1}{a}t_3\,.
\end{equation}
Thus, close to the singular endpoint, both fields become aligned along the Cartan generator $t_3$.

Finally, in the formal limit $a\to0$, the metric approaches flat Euclidean space, and the Abelian, Yang--Mills, and Higgs field strengths become trivial. However, the dilaton behaves as
\begin{equation}
    \phi\sim -\log\left(\frac{r}{a}\right)\,.
\end{equation}
and therefore diverges. Consequently, although the geometry becomes flat in this limit, the full field configuration does not admit a smooth $a\to0$ limit.

This solution provides an example of a partially composite stealth system, in which the gravitational backreaction of only a subset of the matter fields is mutually screened, while the remaining sectors continue to source the geometry. As a consequence, the background is no longer flat, even though part of the matter content remains gravitationally invisible.

\section{Conclusions and Perspectives}\label{sec:discussion}
Stealth configurations are of particular interest because they provide explicit examples of nontrivial matter fields that leave the background geometry completely unaffected, effectively rendering part of the matter content gravitationally invisible. They therefore offer a useful setting in which to investigate how nontrivial field configurations can evade gravitational backreaction. Among the simplest examples are Yang--Mills instantons, defined as finite-action solutions of the Euclidean Yang--Mills equations interpolating between homotopically inequivalent vacuum sectors. Remarkably, despite the highly nonlinear character of the theory, these configurations satisfy first-order (anti-)self-duality equations. These conditions automatically imply the full second-order Yang--Mills equations and, at the same time, force the energy-momentum tensor to vanish identically, thereby eliminating any gravitational backreaction.

In this work, we considered four-dimensional Euclidean Einstein--Yang--Mills theory and focused on Yang--Mills instantons supported on gravitational-instanton backgrounds, in particular on the (anti-)self-dual Taub--NUT and Eguchi--Hanson geometries. After performing a dimensional reduction along the compact coordinate generating the $U(1)$ fibration of these spaces, the resulting effective theory is a three-dimensional Einstein--Yang--Mills--Higgs--dilaton system coupled to an Abelian gauge field. Despite the strong nonlinear interactions among these fields, exact solutions can be obtained directly from the dimensional reduction once the lower-dimensional degrees of freedom are properly identified.

The reduction of Yang--Mills instantons on Taub--NUT gives rise to a composite stealth configuration on three-dimensional flat Euclidean space. Remarkably, the nontrivial reduced fields interact in such a way that their individual contributions to the energy-momentum tensor cancel pairwise, yielding a vanishing total backreaction. In this case, both the Abelian dilaton--Maxwell sector and the non-Abelian Yang--Mills--Higgs sector are mutually screened. By contrast, the dimensional reduction of instantons on the Eguchi--Hanson background leads to a partially composite stealth configuration. In that case, only the non-Abelian Yang--Mills--Higgs sector exhibits mutual cancellation of its gravitational backreaction, whereas the dilaton and Abelian gauge field remain gravitationally active and deform the reduced geometry away from flat space.

In addition, the dimensional reduction of the four-dimensional (anti-)self-duality equations gives rise to a set of first-order BPS-type relations among the fields of the reduced theory. These equations provide a direct lower-dimensional imprint of the (anti-)self-dual structure of the parent instanton and explain the screening mechanism responsible for the cancellation of the corresponding energy-momentum tensors.

The present construction therefore suggests a systematic mechanism for generating composite or partially composite stealth configurations in lower-dimensional theories starting from matter sectors that are already gravitationally invisible in the higher-dimensional seed theory. In this sense, dimensional reduction does not merely reorganize the field content, but can decompose a higher-dimensional stealth sector into several lower-dimensional interacting sectors whose backreactions either cancel completely or only partially.

Several natural extensions of the present work deserve further investigation. One possibility is to include a cosmological constant in the higher-dimensional Einstein--Yang--Mills theory. Such systems admit instantonic configurations on gravitational-instanton backgrounds such as $\mathbb{CP}^2$~\cite{Kim:2000mg}. Upon dimensional reduction, the cosmological constant is expected to induce a Liouville-type potential for the dilaton~\cite{Charmousis:2009xr}. It would be interesting to determine how this additional interaction modifies the gravitational screening mechanism and whether composite stealth configurations can persist once the dilaton energy-momentum tensor receives a contribution from the induced potential. Another natural direction is the extension to higher dimensions. Although the conventional notion of self-duality for two-form Yang--Mills field strengths is special to four dimensions, analogous first-order duality conditions can be formulated for higher-rank $p$-form fields in suitable dimensions. It would therefore be worthwhile to investigate whether higher-dimensional self-dual or generalized self-dual matter sectors can similarly descend, under dimensional reduction, into composite stealth configurations in lower-dimensional effective theories. We leave these questions for future work.

\begin{acknowledgments}
The author is grateful to Daniel Flores-Alfonso for his insightful comments on the manuscript, as well as for valuable discussions and suggestions. He also thanks Gustavo Aldaz and Eloy Ay\'on-Beato for their helpful comments and stimulating discussions. The author further acknowledges previous collaborations with Fabrizio Canfora and Crist\'obal Corral, from which several ideas underlying the present work originated.
\end{acknowledgments}

\bibliography{biblio.bib}

\end{document}